\documentclass[conference]{IEEEtran}
\IEEEoverridecommandlockouts
\usepackage{cite}
\usepackage{hyperref}
\usepackage{amsmath,amssymb,amsfonts}
\usepackage{algorithmic}
\usepackage{graphicx}
\usepackage{textcomp}
\usepackage{xcolor}
\usepackage{url}

\usepackage{afterpage} 
\usepackage{makecell} 
\usepackage[a4paper, total={184mm,239mm}]{geometry}
\def\BibTeX{{\rm B\kern-.05em{\sc i\kern-.025em b}\kern-.08em
    T\kern-.1667em\lower.7ex\hbox{E}\kern-.125emX}}

\newcommand{\chao}[1]{\textcolor{black}{#1}}
\newcommand{\Ye}[1]{\textcolor{black}{#1}}
\begin{document}


\title{CD-PIM: A High-Bandwidth and Compute-Efficient LPDDR5-Based PIM for Low-Batch LLM Acceleration on Edge-Device}

\renewcommand{\arraystretch}{1.8} 

\author{
     Ye Lin\textsuperscript{1}, 
     Chao Fang\textsuperscript{1}, 
     Xiaoyong Song\textsuperscript{2}, 
     Qi Wu\textsuperscript{1}, 
     Anying Jiang\textsuperscript{1}, 
     Yichuan Bai\textsuperscript{1},
     Li Du\textsuperscript{1,$\dagger$} \\
	\IEEEauthorblockA{
		\textsuperscript{1}School of Electronic Science and Engineering, Nanjing University, China~~~\textsuperscript{2}China Mobile Research Institute, China \\
        \small Email:
		\{yelin, fantasysee, qiwu, jay, yicbai\}@smail.nju.edu.cn, songxiaoyong@chinamobile.com, ldu@nju.edu.cn
    }
}
\maketitle

\begin{abstract}

\chao{Edge deployment of low-batch large language models (LLMs) faces critical memory bandwidth bottlenecks when executing memory-intensive general matrix-vector multiplications (GEMV) operations. While digital processing-in-memory (PIM) architectures promise to accelerate GEMV operations, existing PIM-equipped edge devices still suffer from three key limitations: limited bandwidth improvement, component under-utilization in mixed workloads, and low compute capacity of computing units (CUs).}
\chao{In this paper, we propose CD-PIM to address these challenges through three key innovations.}
\chao{First, we introduce a high-bandwidth compute-efficient mode (HBCEM) that enhances bandwidth by dividing each bank into four pseudo-banks through segmented global bitlines. Second, we propose a low-batch interleaving mode (LBIM) to improve component utilization by overlapping GEMV operations with GEMM operations. Third, we design a compute-efficient CU that performs enhanced GEMV operations in a pipelined manner by serially feeding weight data into the computing core.}
Forth, we adopt a column-wise mapping for the key-cache matrix and row-wise mapping for the value-cache matrix, which fully utilizes CU resources.
\chao{Our evaluation shows that compared to a GPU-only baseline and state-of-the-art PIM designs, our CD-PIM achieves 11.42$\times$ and 4.25$\times$ speedup on average within a single batch in HBCEM mode, respectively.}
Moreover, for low-batch sizes, the CD-PIM achieves an average speedup of 1.12$\times$ in LBIM compared to HBCEM. 
\end{abstract}

\let\thefootnote\relax\footnotetext{$^\dagger$Corresponding author. This work was funded in part by the National Key Research and Development Program of China under Grant 2022YFB4400900, in part by the Strategic Industries and Key Technologies Project of Jiangsu Province under Grant BE2023020-3, in part by the Basic Research Program of Jiangsu Province under Grant BK20243042, and in part by the Nanjing University-China Mobile Communications Group Co., Ltd. Joint Institute.}

\section{Introduction} \label{sec:intro}
Transformer-based large language models (LLMs)\cite{LLM} have demonstrated exceptional capabilities across a wide range of applications, such as question answering\cite{Question} and code completion\cite{Code}. Compared to cloud services\cite{Cloud}, which handle large-batch requests requiring high compute capacity, on-device LLM inference\cite{Fold} typically processes low-batch requests and demands high bandwidth to achieve low-latency interactions in real-time applications\cite{application}.

The inference process in LLMs is typically divided into two stages: the prefill stage and the decode stage\cite{DATELLM1}. In the prefill stage, the LLM processes the entire input token sequence in a single step, primarily relying on compute-intensive general matrix-matrix multiplication (GEMM) operations \cite{DATELLM2}. In contrast, the decode stage generates one token at a time in an autoregressive manner, which relies on memory-intensive general matrix-vector multiplication (GEMV) operations \cite{DATELLM3}. However, the limited memory bandwidth of the edge device leads to increased decode latency~\cite{ASTER}.

To mitigate memory bandwidth bottlenecks in the decode stage, digital processing-in-memory (PIM) architectures are increasingly adopted as hardware accelerators. Based on the placement of computing unit (CU), PIM designs can be categorized into subarray-level\cite{subarray}, bank-level\cite{BANK}, bankgroup (BG)-level\cite{AttAcc}, and I/O-level\cite{IO-level} architectures. Among these, bank-level PIM offers a favorable trade-off between area cost and internal memory bandwidth by exploiting bank-level parallelism, where CUs are integrated within the memory bank\Ye{\cite{Pipe}}. This approach is prevalent in recently released production-grade PIM devices\cite{SK}. In such architectures, PIM‑equipped processors perform the GEMM operations on the host processor while offloading the GEMV operations to PIM, thereby improving the overall performance of LLM inference~\cite{PIMoE, PLAIN, P3LLM, UMPIM}.

\Ye{To achieve efficient on-device inference, the current PIM designs mainly face three challenges:} 
(1) The previous work AttAcc\cite{AttAcc} leverages the large number of high-bandwidth memory (HBM) banks to achieve an internal bandwidth of 242 TB/s for cloud services. However, edge devices have far fewer memory banks, making existing PIM architectures essential for efficiently accelerating LLM inference on the edge. This resource limitation prevents on-device inference from meeting strict low-latency requirements. In addition, existing PIM architectures employ CU with relatively low compute capacity for GEMV operations, as they operate at an internal memory clock frequency lower than the maximum supported digital clock frequency\cite{Pipe}, and only one CU is integrated per bank to perform GEMV operations \cite{CU}.
(2) Current PIM architectures operate in a blocked mode, where the processor and PIM cannot execute simultaneously, which limits the efficiency of PIM. In particular, for compute‑intensive workloads with long input sequences and short output sequences, the prefill latency on the processor becomes the dominant factor in the end‑to‑end inference latency. A straightforward solution would be to use PIM as a stand-alone accelerator with dedicated memory, enabling simultaneous execution of GEMV operations in the PIM and GEMM operations in the processor \cite{Dedicated1} \cite{Dedicated2}. However, this approach is impractical for edge devices due to constrained memory resources. Therefore, PIM must be integrated into the existing memory system to enable concurrent GEMV and GEMM execution;
(3) In the attention layer, fixed mapping of key (K)-cache matrix and value (V)-cache matrix leads to imbalanced CUs resource utilization due to differing computation flows \cite{AttAcc}. In particular, if the K-cache matrix (\(\textup{H}_{\textup{dim}}\), \(\textup{L}\)) is partitioned row-wise, where \(\textup{H}_{\textup{dim}}\) denotes the embedding dimension and \(\textup{L}\) represents the number of tokens accumulated from the prefill stage up to the current decoding step, the appended (\(\textup{H}_{\textup{dim}}\), 1) column vector is processed by only one CU, leading to imbalanced PIM CUs utilization. However, for the V-cache matrix, row-wise data mapping can fully utilize CU resources.

To address this challenge, we propose cross-division-PIM (CD-PIM), a high-bandwidth and compute-efficient PIM accelerator optimized for low-batch LLM inference on edge devices. The CD-PIM architecture partitions each memory bank into four pseudo-banks (Pbanks), thereby improving internal bandwidth with minor changes to routing and peripheral circuitry. Furthermore, it seamlessly integrates PIM into the existing memory systems and supports two operation modes: high-bandwidth compute-efficient mode (HBCEM) and low-batch interleaving mode (LBIM). In HBCEM, CD-PIM maximizes internal memory bandwidth by simultaneously activating four Pbanks to accelerate GEMV operations, whereas in LBIM, it overlaps the GEMV latency with GEMM latency to reduce end-to-end inference latency. Furthermore, a compute-efficient CU is designed to perform GEMV operations in a pipelined manner.

The main contributions of our work are as follows:
\begin{itemize}

    \item \chao{For challenge (1)}, we propose a high-bandwidth compute-efficient LPDDR5-based CD-PIM architecture that fully accelerates GEMV operations in HBCEM for memory-intensive workloads with short input sequences and long output sequences. The CD-PIM divides each bank into four Pbanks by segmenting the global bitline (GBL), achieving a 4 $\times$ improvement in memory bandwidth. A lightweight high compute CU is designed to efficiently perform GEMV operations in a pipelined manner by serially feeding weight data from the separated buffers into the computing core, enabling higher compute capacity. (Sec.~\ref{subsec:cd_pim})

    \item \chao{For challenge (2)}, we overlap the GEMV latency in PIM with GEMM latency in processor to reduce end-to-end inference latency in LBIM for compute-intensive workload by activating two Pbanks for GEMV operations and another two Pbanks for GEMM operations, enabling the simultaneous execution of PIM and the processor. (Sec.~\ref{subsec:dual_mode})

    \item \chao{For challenge (3)}, we adopt column-wise mapping for the K-cache matrix and row-wise mapping for the V-cache matrix, which fully utilizes CU resources through inner-product and outer-product operations. (Sec.~\ref{subsec:mapping})

\end{itemize}

We evaluate the proposed architecture on both compute-intensive and memory-intensive workloads using an NVIDIA Jetson AGX Orin 64\,GB \cite{nvidia_jetson_orin} and an Apple iPhone 15 Pro \cite{smith_m3_soc}. Our evaluation demonstrates that compared to a GPU-only baseline and state-of-the-art (SOTA) PIM designs, the CD-PIM achieves an average of 11.42$\times$ and 4.25$\times$ speedup in HBCEM. For compute-intensive workloads with low batch sizes, the CD-PIM achieves an average of 1.12$\times$ speedup in LBIM compared to HBCEM.

\section{Background} \label{sec:intro}
\subsection{\textit{Transformer-based LLM Inference}}
Transformer-based LLM consists of multiple cascaded Transformer decoder layers \cite{Anda}. For each request, inference begins with the prefill stage, in which multiple tokens from the input sequence are processed in parallel to generate the first output token \cite{APTLLM}. The decode stage then generates subsequent tokens iteratively, each time using the previously generated token as input. This process continues until the entire output sequence is produced. Owing to this iterative nature, the input to each decoder layer during the decode stage is a single vector, leading to memory-intensive GEMV operations, in contrast to the GEMM operations performed in the prefill stage. Since the prefill stage is executed only once per request, whereas the decode stage must be repeated hundreds of times, the decode latency dominates the end-to-end inference latency.

\subsection{\textit{Bank-level PIM-equipped Accelerators}}
Bank-level PIM‑equipped LLM accelerators typically place the CU within the bank to exploit higher internal memory bandwidth than the external memory interface. 
To overcome the limited bandwidth improvements on device, FOLD-PIM \cite{Fold} divides the GBL into upper and lower segments and employs a single CU operating at twice the internal memory clock frequency. Similarly, Pipe-PIM \cite{Pipe} divides the GBL into left and right segments and employs two CUs, each operating at the internal memory clock frequency. Although these PIM architectures highly efficiently accelerate the GEMV operation, they operate in blocked mode, thereby limiting PIM utilization.

\afterpage{%
\begin{figure}[t]
    \centering
    \includegraphics[width=0.5\textwidth]{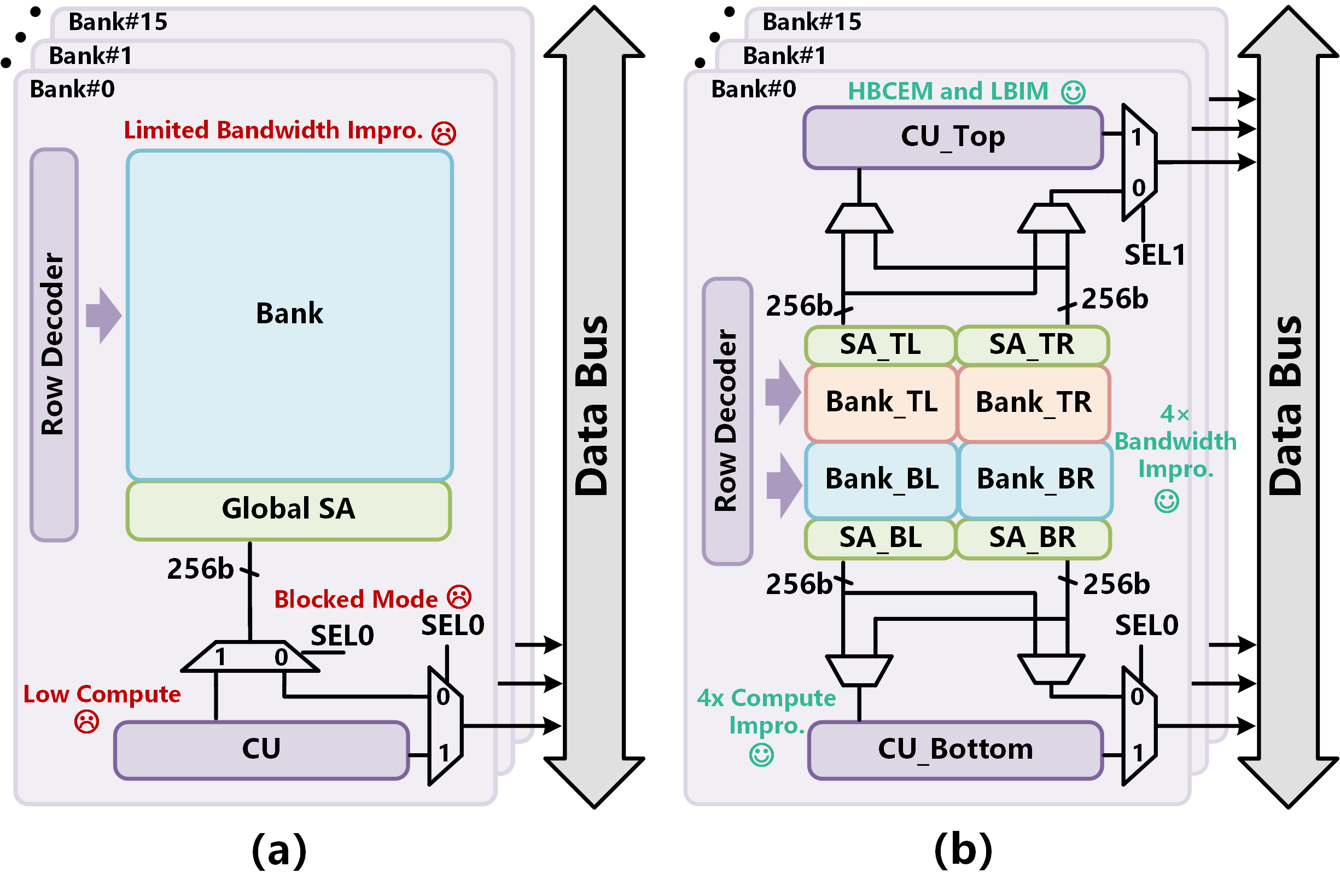}
    \caption{(a) Conventional PIM architecture with a single CU; (b) CD-PIM architecture with two CUs for performing the GEMV operation.}
    \label{fig:pim_cmp}
\end{figure}
}

\afterpage{%
\begin{figure}[t]
    \centering
    \includegraphics[width=0.5\textwidth]{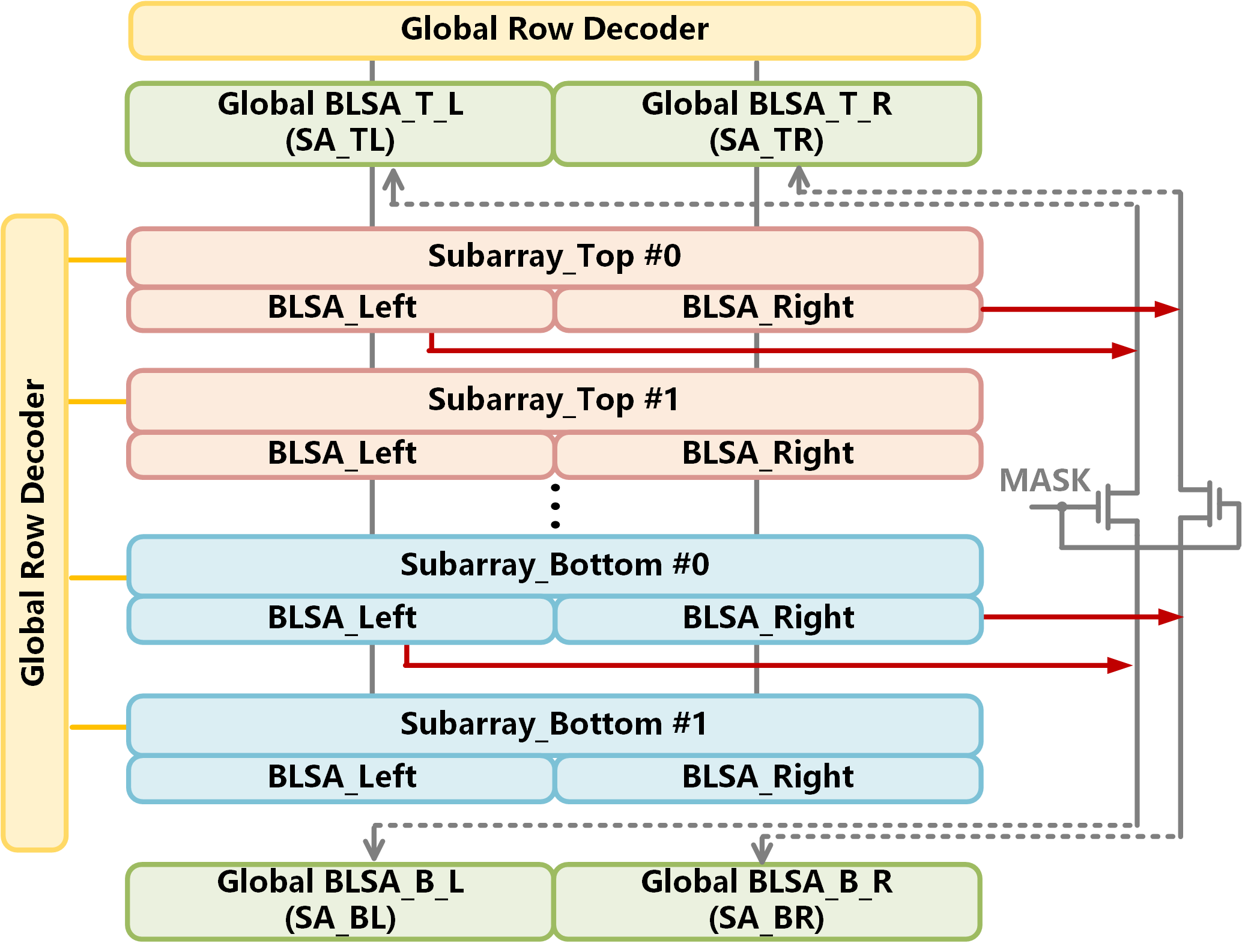}
    \caption{Details of the CD-PIM architecture, where each bank is divided into four Pbanks by separating the GBL and isolation transistors.}
    \label{fig:pim_arch}
\end{figure}
}

A digital PIM integrates CU for performing GEMV operations and a vector buffer for storing and reusing the input vector and partial sums at the bank’s periphery. In bank‑level digital PIM, the CU is placed within the bank region. However, the vector buffer may reside in the bank region \cite{Fold}, the BG region \cite{AttAcc}, or the I/O region \cite{IO-level}. Locating the vector buffer farther from the CU increases data transfer energy. To minimize transfer energy, this work places the vector buffer adjacent to the computing unit.

On-device LLM inference services typically exhibit two key characteristics: handling only a small number of requests concurrently and requiring low-latency interaction. For example, AIOS \cite{AIOS} is presented as an LLM‑embedded management system that supports six interfaces. However, current PIM architectures operate in blocked mode, preventing concurrent operation of PIM and processor computations \cite{Block}. This serialization inherently under-utilizes resources. In particular, for low-batch scenarios, the blocked computation mode leads to longer inference latency, constraining PIM adoption in edge devices. Although prior work\cite{NEU} enables PIM–NPU co-processing via dual row buffer banks, this approach incurs considerable area overhead.

Compared with HBM‑based PIM \cite{AttAcc}, which integrates four dies per rank and over 128 banks per die for cloud devices, LPDDR5‑PIM provides only a limited number of banks, restricting bank‑level parallelism to enhance internal memory bandwidth. Therefore, expanding the internal memory bandwidth is crucial for efficiently accelerating GEMV operations on edge devices. Moreover, the variable workload distribution \cite{H2} and the diversity of compute and memory resources across edge devices \cite{FACIL} largely determine the proportion of time‑to‑first‑token (TTFT) in the overall inference latency. In particular, for compute‑intensive workloads with long input sequences and short output sequences, constrained compute resources cause TTFT to dominate the end‑to‑end inference latency.



In conventional DRAM dies, the bandwidth of the external memory interface exceeds the internal clock frequency due to the burst-transfer scheme \cite{GDDR7}. However, prior bank‑level digital PIM designs \cite{DH_PIM} execute GEMV operations at the internal clock frequency, thereby limiting CU compute capacity. This reduced compute capacity prevents such PIM architectures from fully accelerating low‑batch LLM inference.

\begin{table}[t]
\centering
\small
\caption{Two Edge Device Specifications}
\label{tab:device_memory}
\begin{tabular}{|c|c|c|c|}
\hline
\textbf{Device} & \textbf{DRAM} & \makecell{\textbf{Throughput} \\ \textbf{(TFLOPS)}} & \makecell{\textbf{Bandwidth} \\ \textbf{(GB/s)}} \\
\hline
Jetson AGX Orin & LPDDR5 & 42.5 & 204.8 \\
\hline
iPhone 15 Pro          & LPDDR5 & 4.29 & 51.2 \\
\hline
\end{tabular}
\end{table}

\section{CD-PIM: High-Bandwidth Compute-Efficient PIM Architecture } \label{sec:HBCE}
In this section, we introduce CD-PIM, a high-bandwidth compute-efficient PIM architecture that accelerates GEMV operations by exploiting bank-level parallelism at low implementation cost. CD-PIM leverages its high memory bandwidth to accelerate the decode stage in HBCEM. Moreover, to support low‑batch, compute‑intensive workloads, \Ye{a CD‑PIM‑equipped processor overlaps decode latency by simultaneously executing CD‑PIM for the decode stage and the processor for the prefill stage in LBIM.} Notably, both the input and weight data in this paper are represented with 8-bit precision, which does not lead to any noticeable degradation in LLM inference accuracy. In addition, we introduce three new instructions to support PIM operation.

\subsection{\textit{The Proposed CD-PIM}} \label{subsec:cd_pim}

Compared with conventional DRAM-PIM architectures, as shown in Fig.~\ref{fig:pim_cmp}(a), CD-PIM integrates additional CU directly into each bank. Fig.~\ref{fig:pim_cmp}(b) illustrates the overall architecture of the CD-PIM with two CUs for accelerating GEMV operation. The top and bottom CU perform INT8 multiplication-and-accumulation (MAC) operations in a pipelined manner by receiving weight data from banks through the left and right global BLSAs (SA\_TL, SA\_TR, SA\_BL, SA\_BR). Each CU can handle a 32\,B data width per compute cycle. To enhance compute capacity, the CU is designed to operate at 400\,MHz, which is twice the internal clock frequency of LPDDR5\cite{Fold}. As a result, the CU can pipelined process (64,\,1) vector data from both the left-SA and right-SA within a single memory cycle. Each CU is equipped with a 64\,B input buffer for storing the input vector and a 128\,B output buffer for storing partial sums.
By employing two pipelined CUs in each bank, CD‑PIM provides a $2\times$ computational advantage over DH‑PIM\cite{DH_PIM} equipped with the same DRAM die. In addition, owing to the dedicated mapping strategies for the K-cache and V-cache matrix, the CU is capable of supporting both inner-product and outer-product operations.

Fig.~\ref{fig:pim_arch} illustrates the details of the CD‑PIM architecture.
First, each bank is divided into a left-bank and a right-bank 
by equally splitting the global bitline (GBL), the bitline sense amplifier (BLSA), and the global BLSA into left-GBL/right-GBL, BLSA-left/BLSA-right, and global BLSA-left/global BLSA-right, respectively. 
Owing to this separation, the left-bank and right-bank can be accessed simultaneously by the column decoder. Furthermore, each left-GBL and right-GBL is further partitioned into an upper-GBL and a lower-GBL through isolation transistors, which divides both the left-bank and the right-bank into an upper-bank and a lower-bank. These upper and lower banks can be accessed simultaneously by the row decoder.  As a result, CD-PIM partitions each bank into four Pbanks (Bank\_TL, Bank\_TR, Bank\_BL, Bank\_BR), all of which can be activated concurrently to perform GEMV operations, therefore achieving a higher memory bandwidth compared to FOLD-PIM \cite{Fold}.

\begin{table}[t]
\centering
\renewcommand{\arraystretch}{1}  
\caption{Command Selection Table}
\begin{tabular}{|>{\centering\arraybackslash}m{3cm}|>{\centering\arraybackslash}m{2cm}|>{\centering\arraybackslash}m{2cm}|}
\hline

\textbf{Command Description} & \textbf{SEL0} & \textbf{SEL1} \\
\hline
PIM\_MAC\_FM & 1 & 1 \\
\hline
MACT\_LDB & 0 & 1 \\
\hline
MACB\_LDT & 1 & 0 \\
\hline
\end{tabular}
\label{tab:command}
\end{table}

\subsection{\textit{HBCEM and LBIM}} \label{subsec:dual_mode}
Table~\ref{tab:command} summarizes the three \(\textup{PIM}\) instructions by controlling the \(\textup{SEL0}\) and \(\textup{SEL1}\) signals. The \(\textup{PIM\_MAC\_FM}\) instruction activates four \(\textup{Pbanks}\) simultaneously to fully accelerate the \(\textup{GEMV}\) operation. 
The \(\textup{MACT\_LDB}\) instruction activates the top \(\textup{CU}\) to execute the \(\textup{GEMV}\) operation, whilethe processor performs the GEMM operation by accessing data from the bottom bank. Similarly, the \(\textup{MACB\_LDT}\) instruction activates the bottom \(\textup{CU}\) to execute the \(\textup{GEMV}\) operation while accessing data from the top bank.
Therefore, by simultaneously performing GEMV in PIM and GEMM in the processor, the GEMV latency is overlapped with the GEMM latency. 

For conventional edge devices, the prefill and decode stages of LLM inference are executed on the processor. However, for memory-intensive workloads, the limited memory bandwidth causes significant inference latency, as shown in Fig.~\ref{fig:pim_scheduling}(a). For (Lin, Lout) = (128, 2048) with LLAMA-1B on the NVIDIA Jetson AGX Orin 64 GB, the end-to-end inference latency reaches 35.7s, while the processor compute utilization is 85\%. To accelerate decode stage, we offload the decode stage to PIM in HBCEM, activating four Pbanks to fully accelerate the GEMV operation using \texttt{PIM\_MAC\_FM} instruction as illustrated in Fig.~\ref{fig:pim_scheduling}(b). By exploiting the high internal memory bandwidth of PIM, the decode stage latency is reduced by 90.2\% compared with processor execution, leading to a total inference latency of 3.53s.

For compute-intensive workloads, although \(\textup{PIM}\) accelerates GEMV operation, the TTFT dominates the end-to-end inference latency, particularly on low compute capacity edge devices, as shown in Fig.~\ref{fig:pim_scheduling}(b). For (Lin, Lout) = (2048, 128) with \(\textup{LLAMA-13B}\), offloading the decode stage to \(\textup{PIM}\) causes the \(\textup{TTFT}\) portion of the latency to increase from \(53.7\%\) on the \(\textup{NVIDIA Jetson AGX Orin 64 GB}\) to \(65.2\%\) on the \(\textup{Apple iPhone 15 Pro}\). 
To address this issue, CD-PIM executes GEMV and GEMM operations concurrently, activating two Pbanks for the GEMV operations and another two Pbanks for the GEMM operations in LBIM, using the \texttt{MACT\_LDB} and \texttt{MACB\_LDT} instructions.
Although the \(\textup{PIM}\) compute capacity in LBIM is reduced by half compared to HBCEM, the GEMV latency is effectively overlapped with GEMM latency, as illustrated in Fig.~\ref{fig:pim_scheduling}(c).

\begin{figure}[t]
    \centering
    \includegraphics[width=0.5\textwidth]{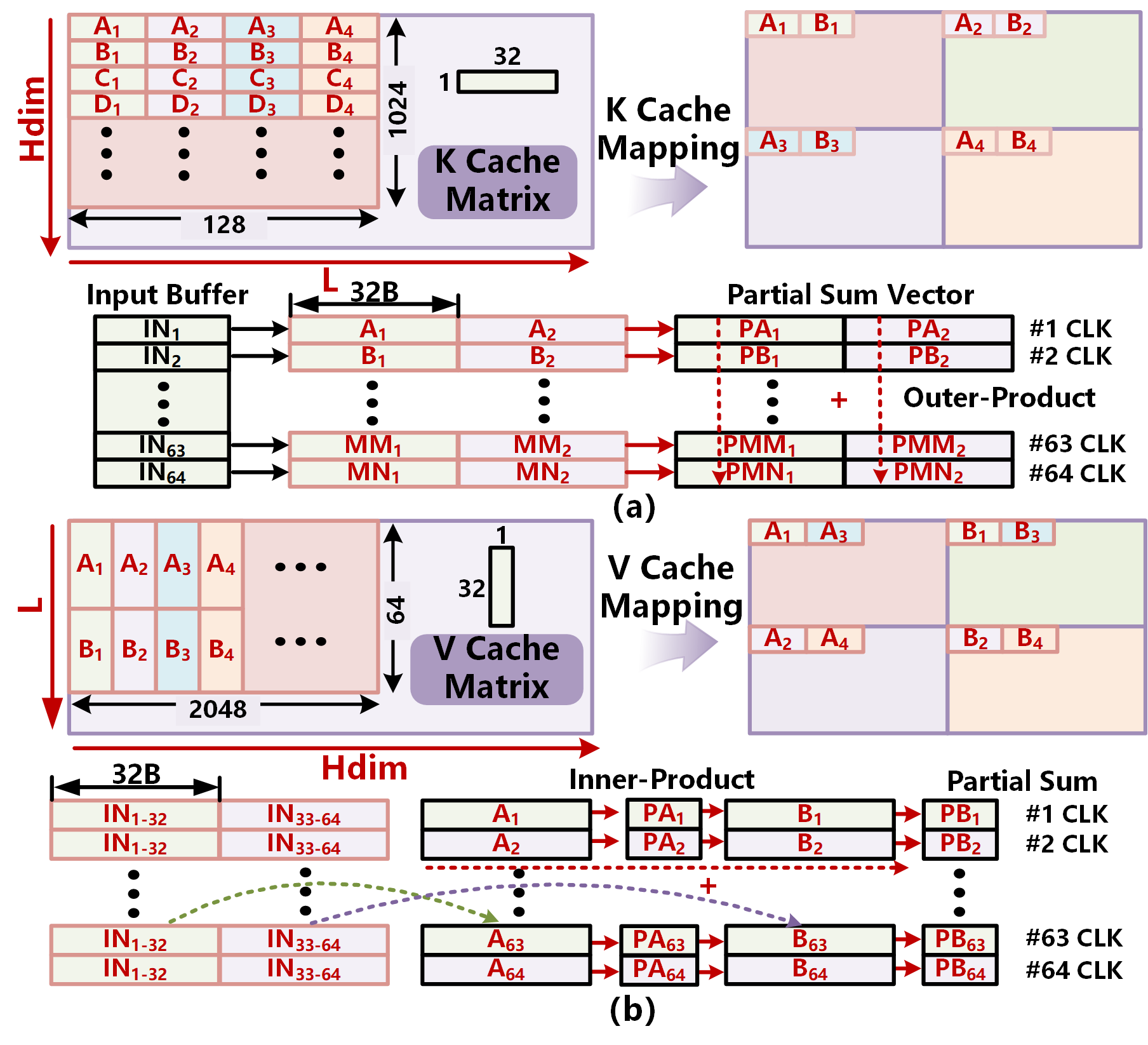}
    \caption{Detailed data mapping of the CD-PIM: (a) K-cache matrix data mapping and outer-product compute flow; (b) V-cache matrix data mapping and inner-product compute flow.}
    \label{fig:pim_mapping}
\end{figure}

\subsection{\textit{Data Mapping}} \label{subsec:mapping}
In the decode stage, the computation flow is composed of the QKV generation layer, the attention layer, and the feed-forward network (FFN). 
First, the input token is fed into the QKV generation layer to derive the query (\(\textup{Q}\)), \(\textup{K}\), and \(\textup{V}\) vectors of size \((1, \textup{H}_{\textup{dim}})\).
The newly generated \(\textup{K}\) and \(\textup{V}\) vectors are concatenated with the  \(\textup{K}\)-cached matrix \((\textup{H}_{\textup{dim}}, \textup{L})\) and \(\textup{V}\)-cached matrix \((\textup{L}, \textup{H}_{\textup{dim}})\).
Subsequently, the \(\textup{Q}\) vector is forwarded to the attention layer, where a GEMV operation is performed with the \(\textup{K}\)-cached matrix. The resulting vector is normalized using the softmax function to produce the attention weight vector \((1, \textup{L})\). 
This weight vector is then multiplied with the \(\textup{V}\)-cached matrix to compute the attention output \((1, \textup{H}_{\textup{dim}})\). 
Finally, the attention output is passed to the FFN layer to generate the next token.

During the attention layer, the data mapping of the \(\textup{K}\textup{V}\)-cached matrix is adjusted according to the compute flow. \textbf{For the \(\textup{K}\)-cache matrix mapping in column-wise, each data chunk corresponds to a vector of size \((1 \times 32)\), which can be accessed by a \(\textup{Pbank}\) in a single burst operation.} At the bank level, each bank is able to access a matrix organized as a \((2 \times 64)\) data chunk. Before the \(\textup{GEMV}\) operation, the \(\textup{Q}\) vector is loaded into the \(\textup{CU}\) input buffer. The \(\textup{Q}\) vector is partitioned into sub-vectors of size \((1 \times 1024)\), which are sequentially transferred to the \(\textup{LPDDR5}\) dies. Each \((1 \times 1024)\) sub-vector is further divided into sixteen \((1 \times 64)\) sub-vectors, corresponding to the \(16\) banks of the \(\textup{LPDDR5}\) device. Within each bank, a \((1 \times 64)\) sub-vector is broadcast to the input buffers of the two \(\textup{CUs}\).  

During the \(\textup{GEMV}\) operation, each \(\textup{CU}\) performs an outer product between the \(1\textup{B}\) input data stored in the input buffer and a \((1 \times 32\textup{B})\) weight vector, producing a \((1 \times 32)\) partial-sum vector. The same input data is then multiplied with another \((1 \times 32\textup{B})\) weight vector to obtain a new \((1 \times 32)\) partial-sum vector. Consequently, within a single internal memory clock cycle, the \(\textup{CU}\) multiplies the \(1\textup{B}\) input data with \((1 \times 64\textup{B})\) weight data, resulting in a \((1 \times 64)\) partial-sum vector. In the following cycle, the produced \((1 \times 64)\) partial-sum vector is accumulated with the results generated in the previous step, as illustrated in Fig.~\ref{fig:pim_mapping}(a).  
For example, the input data \(\textup{IN}_{1}\) is multiplied with the weight vector \(\textup{A}_{1}\) to produce the partial-sum vector \(\textup{PA}_{1}\). The same input data \(\textup{IN}_{1}\) is also multiplied with the weight vector \(\textup{A}_{2}\), yielding the partial-sum vector \(\textup{PA}_{2}\). Next, the input data \(\textup{IN}_{2}\) is multiplied with the weight vector \(\textup{B}_{1}\) and added to \(\textup{PA}_{1}\) to generate the updated partial-sum vector \(\textup{PB}_{1}\). Similarly, \(\textup{IN}_{2}\) is multiplied with the weight vector \(\textup{B}_{2}\) and accumulated with \(\textup{PA}_{2}\), producing the updated partial-sum vector \(\textup{PB}_{2}\).  
Following this compute flow, each bank can perform a \(((1,1) \times (1,128))\) \(\textup{GEMV}\) operation within a single memory internal clock cycle. For each \((1 \times 1024)\) input-vector segment, the input buffer capacity supports the processing of a \((64 \times 128)\) matrix per bank. At the die level, a \((1024 \times 128)\) matrix can be processed. Subsequently, the next \((1 \times 1024)\) input-vector segment is transferred to the \(\textup{DRAM}\) die, enabling continuous computation.

\textbf{For the \(\textup{V}\)-cache matrix mapping in row-wise, three main differences are identified compared to the \(\textup{K}\)-cache matrix, as shown in Fig.~\ref{fig:pim_mapping}(b).} First, each data chunk corresponds to a \((32 \times 1)\) vector, and each bank can access a matrix organized as a \((64 \times 2)\) data chunk, whereas the \(\textup{K}\)-cache matrix accesses a \((2 \times 64)\) data chunk. Second, the attention vector \((1, \textup{L})\) is divided into multiple \((1 \times 64)\) sub-vectors. Each sub-vector is broadcast to all input buffers of the \(\textup{CUs}\). Third, during the \(\textup{GEMV}\) operation, each \(\textup{CU}\) performs an inner product between a \((1 \times 32)\) attention vector \(\textup{IN}_{1-32}\) and a \((32 \times 1)\) weight vector \(\textup{A}_{1}\) to generate a partial sum \(\textup{PA}_{1}\). The subsequent \((1 \times 32\textup{B})\) attention data \(\textup{IN}_{33-64}\) is then multiplied with a \((32 \times 1\textup{B})\) weight vector \(\textup{B}_{1}\) and added to \(\textup{PA}_{1}\) to produce a new partial sum \(\textup{PB}_{1}\).  
Following this computation flow, each bank can execute a \(((1,64) \times (64,2))\) \(\textup{GEMV}\) operation within a single internal clock cycle. For each \((1 \times 64)\) attention vector, the output buffer capacity supports the processing of a \((64 \times 128)\) matrix within a bank, and at the die level, a \((64 \times 2048)\) matrix can be processed. The next \((1 \times 64)\) attention vector is then transferred to the \(\textup{DRAM}\) die for continuous computation.

\section{Evaluation} \label{sec:HBCE}

\subsection{\textit{Methodology}}
We implemented CD-PIM on top of 4GB LPDDR5 by modifying Ramulator2 \cite{Ramulator}, an open-source DRAM simulator. Each LPDDR5 die provides 16 data pins, operating at a rate of 6.4 Gbps per pin. 
\afterpage{%
\begin{figure}[htbp]
    \centering
    \includegraphics[width=0.5\textwidth]{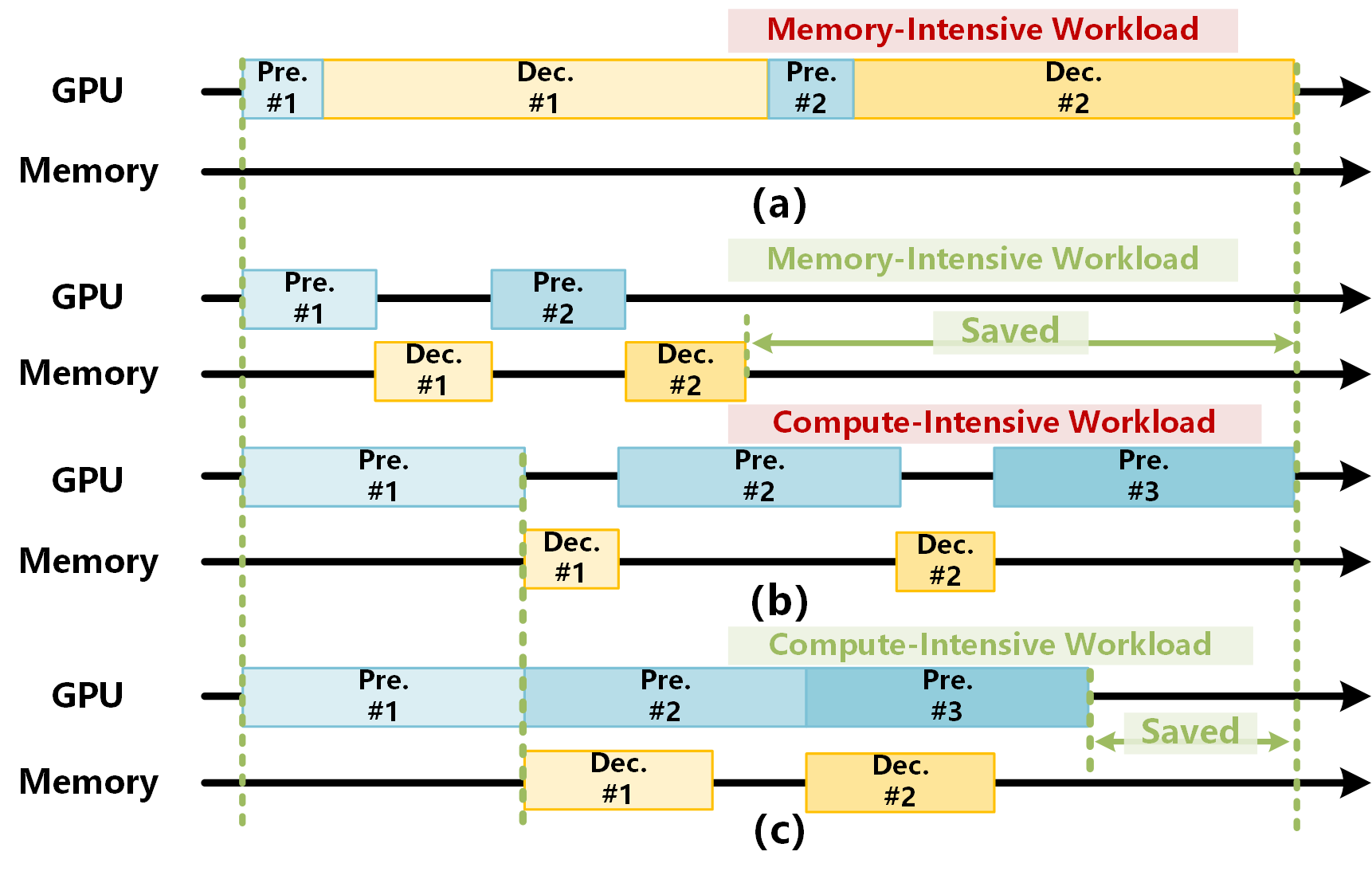}
    \caption{Timing diagrams of LLM inference: (a) GPU with memory-intensive workload; (b) CD-PIM-equipped GPU in HBCEM for memory- and compute-intensive workload; (c) CD-PIM-equipped GPU in LBIM for compute-intensive workload.}
    \label{fig:pim_scheduling}
\end{figure}
}
\Ye{In our experimental setup, we employ the Apple iPhone 15 Pro and the NVIDIA Jetson AGX Orin as edge devices to evaluate the performance of CD-PIM, which are equipped with 4 and 16 LPDDR5 dies, respectively, providing total memory capacities of 16 GB and 64 GB.}
For performance evaluation, we compare CD-PIM with GPU-only baseline and AttAcc\cite{AttAcc} baseline execution on the Apple iPhone 15 Pro and the NVIDIA Jetson AGX Orin. We further assess CD-PIM under batch sizes of one and four, considering both HBCEM and LBIM across different LLM parameters.

\subsection{\textit{Performance Results}}
\begin{figure*}[t]
    \centering
    \includegraphics[width=1\textwidth]{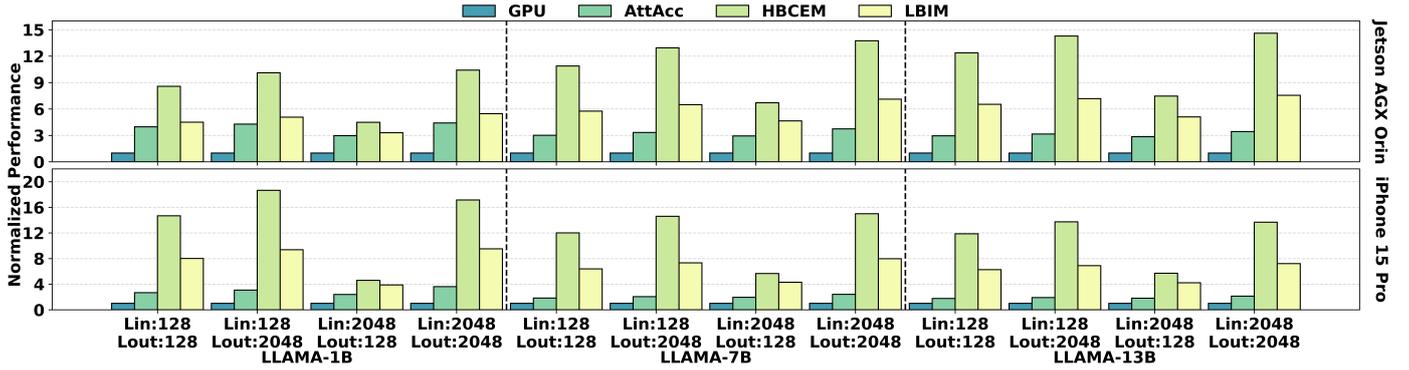}
    \caption{Normalized performance of LLaMA-1B, -7B, and -13B (batch size $=1$) under various Lin and Lout on the NVIDIA Jetson AGX Orin and Apple iPhone~15 Pro.}
    \label{fig:perf_bs_one}
    \Ye{\vspace{-20pt}}
\end{figure*}

Fig.~\ref{fig:perf_bs_one} shows the normalized performance evaluation of the GPU and AttAcc compared with CD-PIM in HBCEM and LBIM under a single-batch scenario. On the NVIDIA Jetson AGX Orin, CD-PIM in HBCEM achieves a speedup of 4.48$\times$--10.51$\times$ for LLAMA-1B, 6.71$\times$--13.74$\times$ for LLAMA-7B, and 7.47$\times$--14.6$\times$ for LLAMA-13B relative to the GPU-only baseline \cite{LLM1B-13B}. These increases result from exploiting the higher internal memory bandwidth of PIM compared to the external memory bandwidth of the GPU. Furthermore, as the model size increases from LLAMA-1B to LLAMA-13B, the decode stage incurs higher latency. By accelerating the GEMV operation with PIM, the performance improvements become more pronounced.

\begin{figure}[t]
    \centering
    \includegraphics[width=0.5\textwidth]{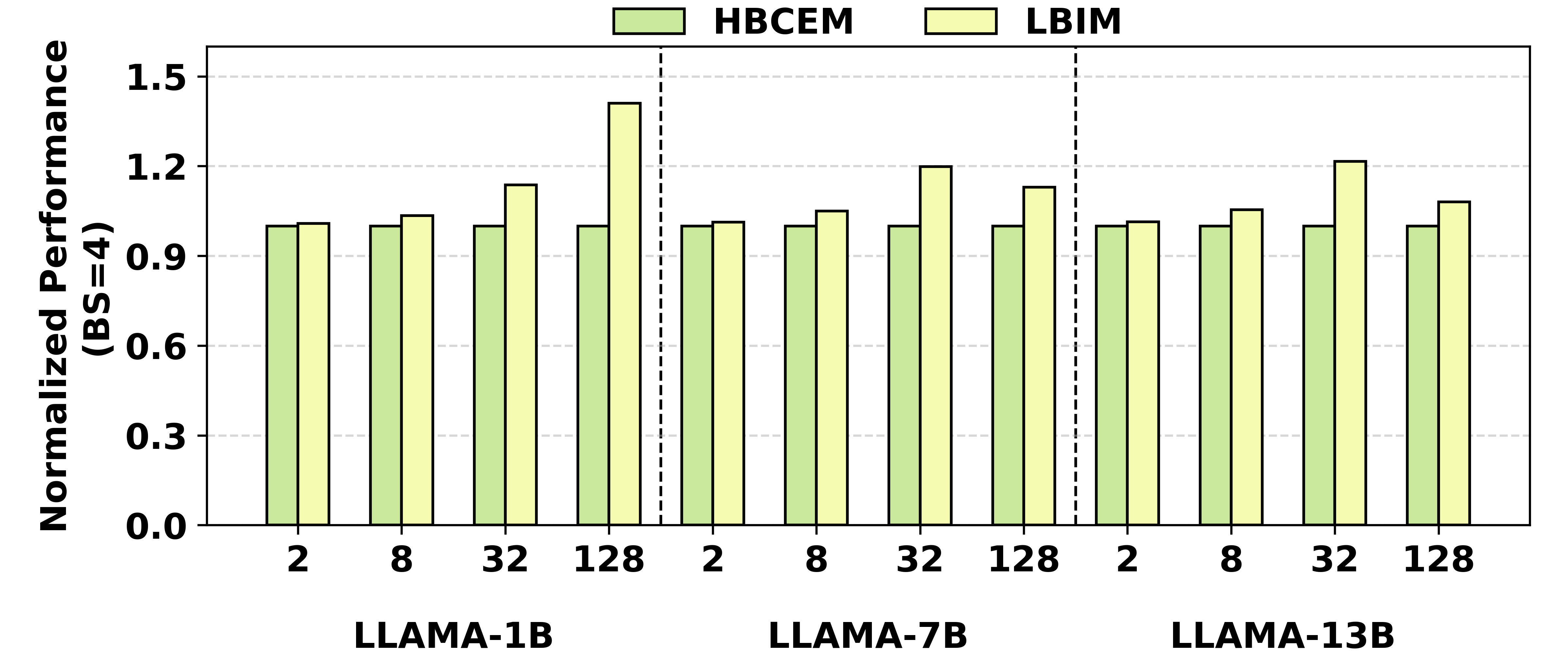}
    \caption{Normalized performance of the NVIDIA Jetson AGX Orin equipped with CD‑PIM under HBCEM and LBIM (batch size $=4$).}
    \label{fig:perf_bs_four_orin}
\end{figure}

\begin{figure}[t]
    \centering
    \includegraphics[width=0.5\textwidth]{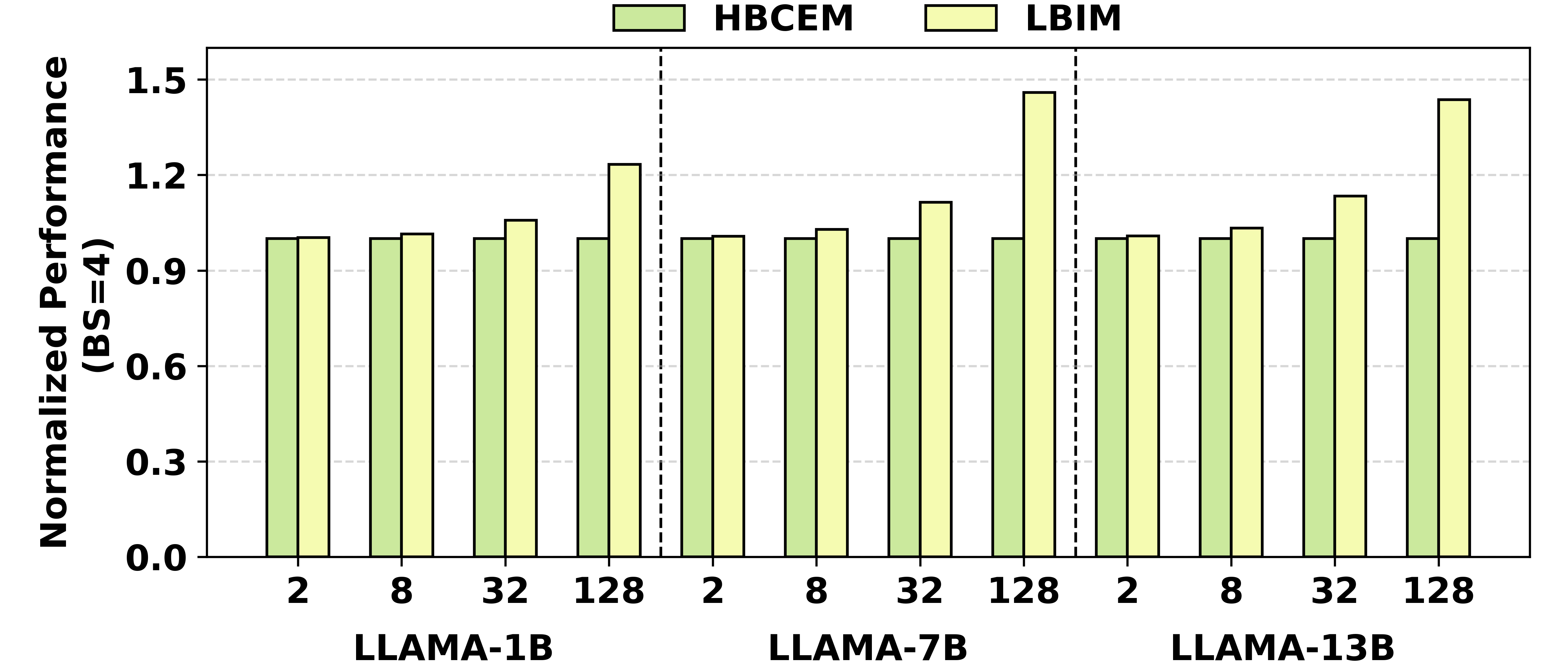}
    \caption{Normalized performance of the Apple iPhone~15 Pro equipped with CD‑PIM under HBCEM and LBIM (batch size $=4$).}
    \label{fig:perf_bs_four_iphone}
    \Ye{\vspace{-15pt}}
\end{figure}

For the Apple iPhone 15 Pro, CD-PIM in HBCEM achieves a higher speedup compared to the NVIDIA Jetson AGX Orin. Owing to the iPhone's lower memory bandwidth relative to the Jetson, the decode stage incurs longer latency, making the workload more memory-intensive. Consequently, accelerating the GEMV operation with PIM yields greater performance improvements on the iPhone. In particular, for the memory-intensive case with (Lin, Lout) = (128, 2048), the speedup increases from 10.1$\times$ on the Jetson AGX Orin to 18.6$\times$ for LLAMA-1B on the iPhone 15 Pro.

However, due to limited compute capacity, CD-PIM in HBCEM is less efficient for compute-intensive workloads compared to memory-intensive workloads. Fig.~\ref{fig:perf_bs_four_orin} presents the performance evaluation of CD-PIM in both HBCEM and LBIM under a four-batch scenario on the NVIDIA Jetson AGX Orin platform, where Lin is fixed at 2048. For LLAMA-1B, when Lout increases from 2 to 128, CD-PIM in LBIM achieves a speedup of 1.01$\times$--1.41$\times$ over HBCEM baseline by overlapping the latency of the decode stage. In contrast, for LLAMA-7B and LLAMA-13B, the performance speedup in LBIM initially grows with Lout from 2 to 32, but then decreases because at Lout = 128, the decode stage latency surpasses the TTFT latency.

For the Apple iPhone~15~Pro, CD-PIM in LBIM achieves a speedup of 1.01$\times$--1.23$\times$ over HBCEM in LLAMA-1B, which is lower than the performance speedup on the NVIDIA Jetson AGX Orin, since for compute-intensive workloads, the reduced compute capacity has a stronger impact than the reduced memory bandwidth. Consequently, the TTFT latency contribution to end-to-end inference latency is higher on the iPhone~15~Pro than on the Jetson AGX Orin. For LLAMA-7B and LLAMA-13B, CD-PIM in LBIM achieves speedups of 1.01$\times$--1.46$\times$ and 1.01$\times$--1.44$\times$ over HBCEM, respectively, when Lout ranges from 2 to 128, because the decode stage latency increases to nearly half of the total inference latency, as shown in Fig.~\ref{fig:perf_bs_four_iphone}.

\begin{figure}[t]
    \centering
    \includegraphics[width=0.5\textwidth]{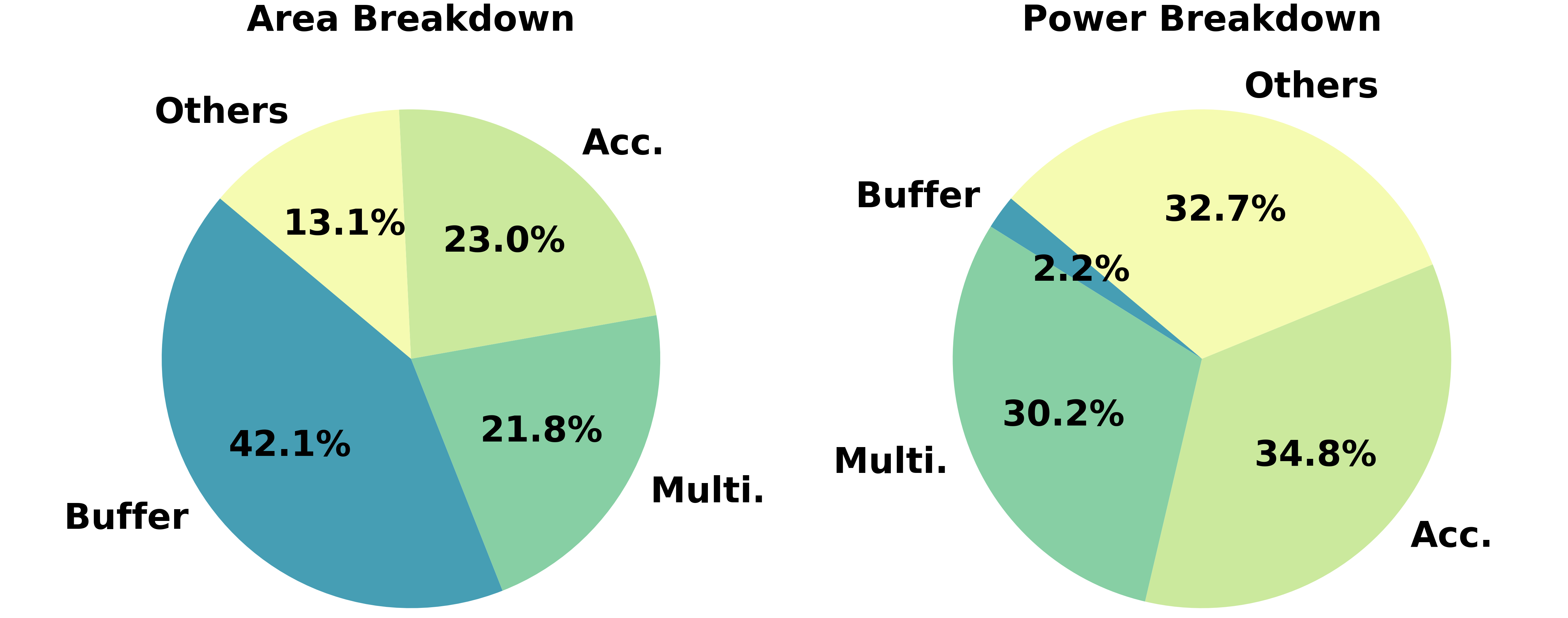}
    \caption{The area and power breakdown of CU.}
    \label{fig:breakdown}
    \Ye{\vspace{-20pt}}
\end{figure}

\subsection{\textit{Area and Power Overhead}}
Compared with CUs that only support inner-products, the proposed CU supports both inner- and outer-products. Fig.~\ref{fig:breakdown} illustrates the area and power breakdown of CU. 
We evaluated the CU area and power using Synopsys Design Compiler, which shows that each PU occupies $14{,}941~\mu\text{m}^2$ and consumes $4.5~\text{mW}$ in TSMC~28\,nm. This corresponds to $0.8\%$ of a 32\,Gb LPDDR5 die area and adds $144~\text{mW}$ in total, confirming the cost efficiency of CD-PIM.

\section{Conclusion} \label{sec:HBCE}
In this paper, we propose CD-PIM to address the inefficiency of GEMV acceleration for memory-intensive workloads and the longer TTFT latency in compute-intensive workloads. The CD‑PIM divides each bank into four Pbanks to enhance internal memory bandwidth and is equipped with two CUs to achieve high computational capacity. For memory-intensive workloads, CD-PIM in HBCEM fully accelerates GEMV operations; for compute-intensive workloads, CD‑PIM in LBIM overlaps the decode latency by executing CD‑PIM for the decode stage and the processor for the prefill stage simultaneously. A high compute capacity CU is designed to execute GEMV in a pipelined manner. We also analyze the area and power overhead of CD-PIM, showing that the proposed architecture delivers significant performance improvements with only minimal hardware cost.

\bibliographystyle{IEEEtran}   
\bibliography{ref}            

\end{document}